\documentclass[a4paper,fleqn]{cas-sc}

\usepackage{color}
\usepackage{amssymb}
\usepackage{amsfonts,amssymb}
\usepackage{amsmath}
\usepackage[numbers]{natbib}

\usepackage{newtxtext}
\usepackage{subcaption}
\usepackage{caption}
\usepackage{booktabs}
\usepackage{amsmath}
\usepackage{graphicx}
\usepackage{booktabs}
\usepackage{algorithm}
\usepackage{algorithmic}
\usepackage{float}
\def\tsc#1{\csdef{#1}{\textsc{\lowercase{#1}}\xspace}}
\tsc{WGM}
\tsc{QE}

\begin{document}
\let\WriteBookmarks\relax
\def\floatpagepagefraction{1}
\def\textpagefraction{.001}
\let\printorcid\relax
\shorttitle{}    

\shortauthors{Z. Yang et~al.}

\title[mode = title]{Atomic Norm Soft Thresholding for Sparse Time-frequency Representation}  



%

\author[inst1,inst2]{Zongyue Yang}





\credit{Conceptualization, Methodology, Visualization, Validation, Software, Writing - original draft, Writing - review $\&$ editing}

\affiliation[inst1]{organization={National Key Lab of Aerospace Power System and Plasma Technology, Xi'an Jiaotong University},
            city={Xi'an},
            postcode={710049}, 
            state={Shannxi},
            country={China}}

\author[inst1, inst2]{Baoqing Ding}
\cormark[1]
\ead{dingbq@xjtu.edu.cn}
\credit{Conceptualization, Resources, Supervision, Writing - review $\&$ editing}

\author[inst1,inst2]{Shibin Wang}
\credit{Resources, Supervision}

\author[inst1,inst2]{Chuang Sun}
\credit{Resources, Supervision}

\author[inst1,inst2]{Xuefeng Chen}
\credit{Funding acquisition,  Resources, Supervision}
\affiliation[inst2]{organization={School of Mechanical Engineering, Xi'an Jiaotong University},
            city={Xi'an},
            postcode={710049}, 
            state={Shannxi},
            country={China}}

\cortext[1]{Corresponding authors}



\begin{abstract}
Time-frequency (TF) representation of non-stationary signals typically requires the effective concentration of energy distribution along the instantaneous frequency (IF) ridge, which exhibits intrinsic sparsity. Inspired by the sparse optimization over continuum via atomic norm, a novel atomic norm soft thresholding for sparse TF representation (AST-STF) method is proposed, which ensures accurate TF localization under the strong duality. Numerical experiments demonstrate that the performance of the proposed method surpasses that of conventional methods.
\end{abstract}


\begin{highlights}
\item The atomic norm soft thresholding method was first introduced into time-frequency analysis.
\item The proposed AST-STF method has good robustness to noise.
\item The proposed AST-STF method exhibits excellent time-frequency energy concentration than the conventional methods.
\end{highlights}

\begin{keywords}
Time-frequency analysis \sep Sparse representation \sep Atomic norm soft thresholding \sep Semidefinite programming
\end{keywords}

\maketitle

\section{Introduction}\label{sec:introduction}
Time-frequency analysis (TFA) provides a framework for analyzing non-stationary signals in TF domain, which is commonplace in practical applications like audio signals \cite{kowalskiSocialSparsityNeighborhood2013}, radar \cite{belouchrani_source_2013} and the vibration of rotating machinery \cite{wang_matching_2014}. Frequently utilized TFA methods include short time Fourier transform (STFT), Stockwell transform (ST) and continuous wavelet transform (CWT). However, these transforms may sometimes fail to achieve ideal TF energy concentration due to the constraints imposed by the Heisenberg uncertainty principle.

Several approaches in the past decades have been proposed to achieve a well-localized TF representation. Among these, the reassignment technique has attracted considerable attention, which effectively enhances the TF energy concentration by estimating the theoretical positions of TF atoms and reallocating the coefficients. Although reassignment effectively improves the performance of TF representation, it is not feasible to reconstruct the signal due to the detriment of linear structure of the original signal. Further studied CWT-based synchrosqueezing transform (SST) reallocates the coefficients in the frequency axis only, which is a special reassignment technique and can preserve the property of signal reconstruction.

Furthermore, TFA methods based on sparse representation are powerful tools to enhance the TF domain sparsity of signals, which have been extensively explored. The core concept of sparse representation is that a linear combination of limited atoms from a certain predefined dictionary can effectively represent a signal. This implies that TFA based on sparse representation can offer more complete information for characterizing varying IF features of non-stationary signals. The typical convex formulation of sparse representation based on $l_1$ norm minimization has garnered widespread attention in research. However, it may lead to performance degradation in the presence of resolution limitations resulting from the discretization of the pre-defined dictionary \cite{chi_sensitivity_2011}. 

In recent years, sparse representation over continuum via atomic norm has been studied \cite{chandrasekaran_convex_2012}. It provides a general convex formulation of sparse representation, and has been widely used in applications such as line spectrum estimation \cite{bhaskar_atomic_2013}, compressed sensing \cite{tang_compressed_2013}, super resolution \cite{chi_harnessing_2020}, and direction of arrival (DOA) \cite{li_off--grid_2016, chi_compressive_2015, zhu_deep_2022}. In addition, its application in TF representation based on a basic pursuit (BP) model illustrates the potential for enhancing the TF energy concentration \cite{kusano_sparse_2021}. However, further study is needed, especially in the realm of signal denoising and reconstruction.

In this paper, we propose a sparse TF representation method based on the atomic norm soft thresholding (AST). Our proposed method transform the sparse TF representation model into a semidefinite programming (SDP) problem, which can be efficiently solved via the Alternating Direction Method of Multipliers (ADMM). Afterwards, the support frequencies in each windowed segment can be accurately localized through the dual polynomial, which returns super-resolution sparse TF representation of signal. Experiments including simulation and practical application confirmed that the proposed method exhibits better performance of signal TF energy concentrating and denoising than the conventional methods. 

The structure of this paper is as follows: In Section \ref{model}, we propose the atomic norm soft thresholding based sparse TF representation model and support frequencies localization method via dual polynomial. Section \ref{sec:simulation} details the simulation analysis, while Section \ref{sec:application} provides application verification using a bat echolocation signal. Section \ref{sec:conclusion} concludes the paper.

{\em Notations:} $\mathbb{R}$ and $\mathbb{C}$ represent the set of real and complex numbers. $\overline {\left(  \cdot  \right)}$, ${\left(  \cdot  \right)^T}$, and ${\left(  \cdot  \right)^ * }$ denote the conjugate, transpose and conjugate transpose, respectively. ${\rm{Tr}}\left( {\bf{X}} \right)$ denotes the trace of matrix $\bf{X}$. $\left\langle {{\bf{x}},{\bf{y}}} \right\rangle  = {{\bf{x}}^T}{\bf{\bar y}}$ is the inner product of $\bf{x}$ and $\bf{y}$. ${\left\|  \cdot  \right\|_F}$ represents the Frobenius norm. ${\bf{Z}}\succeq{0}$ denotes $\bf{Z}$ being positive semidefinite.

\section{Mathematical formulation}\label{model}
\subsection{Atomic norm for line spectrum estimation}
In the application of line spectrum estimation based on sparse representation, a basic metric of sparsity is to exploit the minimum frequency components composing the signal, which is defined as the $l_0$ norm: 
\begin{equation} \label{eq:1}
    {\left\| {\bf{x}} \right\|_0} \buildrel \Delta \over = \inf \left\{ {K:{\bf{x}} = \sum\limits_{k = 0}^K {{c_k}{{\bf{a}}_k}} } \right\},
\end{equation}
where ${{\bf{a}}_k}$ is the discrete sinusoid component, and $c_k$ corresponds the coefficient. Since $l_0$ is a nonconvex and NP-hard problem, $l_1$ is introduced as its convex relaxation and is formulated as non-negative linear combination of atoms from the discrete atomic set.

\begin{figure}[t] 
\centerline{\includegraphics[width=0.4\columnwidth]{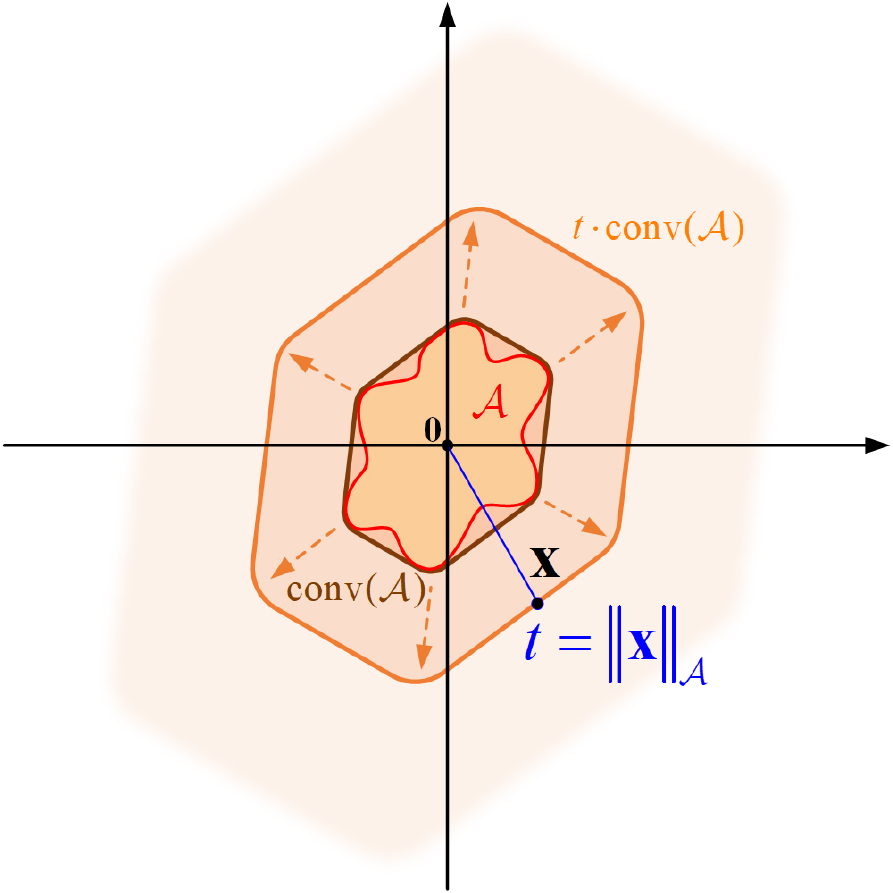}}
\caption{The illustration of an atomic set (in red), its convex hull (in brown) and atomic norm (in blue).}
\label{fig:1}
\end{figure}

As a general framework, atomic norm is formulated by a convex penalty function that specially caters to the facial geometry structure of atomic set $\mathcal{A}$ , as shown in Fig. \ref{fig:1}. It can be denoted as \cite{chandrasekaran_convex_2012}: 
\begin{equation} \label{eq:2}
    {\left\| {\mathbf{x}} \right\|_\mathcal{A}} \triangleq \inf \left\{ {t > 0:{\mathbf{x}} = t \cdot {\text{conv}}(\mathcal{A})} \right\},
\end{equation}
in which $conv(\cal{A})$ corresponds to the convex hull of atomic set $\cal{A}$, and $t$ denotes the scaling factor of $conv(\cal{A})$, i.e., the value of atomic norm. The gauge function of atomic norm is:
\begin{equation} \label{eq:3}
    {\left\| {\bf{x}} \right\|_{\cal A}} = \inf \left\{ {\sum\limits_k {{c_k}:} {\rm{ }}{\bf{x}} = \sum\limits_k {{c_k}{{\bf{a}}_k},{c_k} \ge 0,} \forall {{\bf{a}}_k} \in {\cal A}} \right\},
\end{equation}

Specifically, we consider the signal sampled through Nyquist sampling ${\mathbf{x}} = {\left[ {\begin{array}{*{20}{c}}
  {{x_0}}& \cdots &{{x_{N - 1}}} 
\end{array}} \right]^T} \in {\mathbb{C}^n}$, and assume that $\cal{A}$ is a set of parameterized Fourier bases, which is:
\begin{equation} \label{eq:4}
   {\cal A} = \left\{ {{e^{i2\pi \phi }}{{\left[ {\begin{array}{*{20}{c}}
1&{{e^{i2\pi f}}}& \cdots &{{e^{i2\pi \left( {N - 1} \right)f}}}
\end{array}} \right]}^T},f,\phi  \in \left[ {0,1} \right]} \right\}.
\end{equation}

Since the infinite set $\cal{A}$ exhibits a Vandermonde structure, according to the Proposition II.1 in \cite{tang_compressed_2013}, the atomic norm can be calculated through a computationally feasible semidefinite programming (SDP) formulation:
\begin{equation} \label{eq:5}
    {\left\| {\bf{x}} \right\|_{\cal A}} = \inf \left\{ {\frac{1}{2}\left( {t + {\rm{Tr(Toep}}{\rm{(}}{\bf{u}}{\rm{))}}} \right){\rm{:}}\left[ {\begin{array}{*{20}{c}}
{\rm{Toep}{\rm{(}}{\bf{u}}{\rm{)}}}&{\bf{x}}\\
{{{\bf{x}}^*}}&t
\end{array}} \right] \succeq 0} \right\},
\end{equation}
where ${\mathbf{u}} = {\left[ {\begin{array}{*{20}{c}}
  {{u_1}}&{{u_2}}& \cdots &{{u_N}} 
\end{array}} \right]^T} \in {\mathbb{C}^N}$, and ${\rm{Toep(}}{\bf{u}}{\rm{)}}$ is the Hermitian Toeplitz matrix with its first column being $\bf{u}$. Once we obtain the optimal solution ${\mathbf{\hat u}}$ by solving  (\ref{eq:5}),  the frequency components can be obtained by Vandermonde decomposition:
\begin{equation} \label{eq:6}
    {\rm{Toep(}}{\bf{\hat u}}{\rm{)}} = \sum\limits_{k = 1}^K {\left| {{c_k}} \right|} {\rm{ }}{a_k}a_k^ * ,{a_k} \in {\cal A}.
\end{equation}

\subsection{Atomic norm soft thresholding for sparse TF representation}
The ideal time-frequency distribution is the ultimate goal of time-frequency analysis research. One of the most classic methods for analyzing non-stationary signals is the short time Fourier transform (STFT), which is denoted as:
\begin{equation} \label{eq:7}
    {F_x}(u,f) = \left\langle {x,{w_{u,f}}} \right\rangle  = \int_{ - \infty }^{ + \infty } {x(t)} w(t - u){{\rm{e}}^{ - j2\pi f\left( {t - u} \right)}}{\rm{d}}t,
\end{equation}
where ${w_{u,f}}(t)$ is a Gaussian window function with $u$ and $f$ being its time-shift and modulated frequency. STFT can be seen as the inner product between the signal being truncated by a window and the basis function, which is the Fourier transform of the windowed signal.

Generally, the ideal TF distribution of non-stationary signals is sparse in the TF domain. This implies that sparse representation theory can be connected with TFA by constructing a sparse representation model to enhance the sparsity in TF domain, and thereby improving the energy concentration. Based on the principles above, the atomic norm based TF sparse representation model is formulated as:
\begin{equation} \label{eq:8}
    \mathop {\min }\limits_{\bf{x}} {\left\| {\bf{x}} \right\|_{\cal A}, }{{\text{ s}}{\text{.t}}{\text{. }}}{\bf{y}} = {\bf{Dx}},
\end{equation}
where $\cal{A}$ conforms to the set of parameterized Fourier bases similar to (\ref{eq:4}). The inverse operator ${\mathbf{D}} \in {\mathbb{R}^{N \times (LW)}}$ is the de-window matrix, with which $L$ and $W$ are the length and number of the window, respectively. Considering the measured signal $\bf{y}$ with additive noise, we define the atomic norm soft thresholding method for sparse TF representation (AST-STF) model as: 
\begin{equation} \label{eq:9}
    \arg \mathop {\min }\limits_{\mathbf{x}} \frac{1}{2}\left\| {{\mathbf{y}} - {\mathbf{Dx}}} \right\|_2^2 + \tau {\left\| {\mathbf{x}} \right\|_\mathcal{A}}.
\end{equation}

In (\ref{eq:9}), $\left\| {{\mathbf{y}} - {\mathbf{Dx}}} \right\|_2^2$ is the data fidelity term and $\tau$ is the regularization parameter. The optimum choice of $\tau$ is dependent on the dual atomic norm of noise \cite{tang_compressed_2013}, denoted as:
\begin{equation} \label{eq:10}
    \tau  = \sigma \sqrt {\log (4\pi \log (N)) + \log (N) },
\end{equation}
where $\sigma$ is the standard deviation, $N$ is the length of signal.

The SDP transformation of (\ref{eq:9}) can be denoted as:
\begin{equation} \label{eq:11}
   \begin{gathered}
  \mathop {\min }\limits_{t,{\mathbf{u}},{\mathbf{x}}} \frac{1}{2}\left\| {{\mathbf{y}} - {\mathbf{Dx}}} \right\|_2^2 + \frac{\tau }{2}\left( {t + {\text{Tr}}{\rm(Toep(\mathbf{u}}{\text{))}}} \right), \hfill \\
  {\text{s}}{\text{.t}}{\text{. }}{\mathbf{Z}} = \left[ {\begin{array}{*{20}{c}}
  {{\rm{Toep(}}{\mathbf{u}}{\text{)}}}&{\mathbf{x}} \\ 
  {{{\mathbf{x}}^ * }}&t 
\end{array}} \right],{\mathbf{Z}} \succeq 0. \hfill \\ 
\end{gathered} 
\end{equation}

This SDP problem can be directly solved by SeDuMi, SDPT3 \cite{ttnc_solving_2003} or CVX \cite{wong_cvxbased_2019}. However, since it is nonsmooth convex optimization, the computational cost of these solvers is prohibitive as the signal scale grows. Fortunately, optimization algorithm based on the Alternating Direction Method of Multipliers (ADMM) \cite{boyd_distributed_2010} has been proven to effectively solve the SDP problem. Accordingly, one can dualize the constraint equation through the Augmented Lagrangian:
\begin{equation} \label{eq:12}
    \begin{gathered}
  \mathcal{L}\left( {{\mathbf{u}},{\mathbf{x}},{\mathbf{Z}},{\mathbf{\Lambda }}, t} \right) = \frac{1}{2}\left\| {{\mathbf{y}} - {\mathbf{Dx}}} \right\|_2^2 + \frac{\tau }{2}\left( {t + {\text{Tr(Toep}}{(\mathbf{u}}{\text{))}}} \right) \\ 
   + \left\langle {{\mathbf{\Lambda }},{\mathbf{Z}} - \left[ {\begin{array}{*{20}{c}}
  {{\rm{Toep(}}{\mathbf{u}}{\text{)}}}&{\mathbf{x}} \\ 
  {{{\mathbf{x}}^ * }}&t 
\end{array}} \right]} \right\rangle  + \frac{\rho }{2}\left\| {{\mathbf{Z}} - \left[ {\begin{array}{*{20}{c}}
  {{\rm{Toep(}}{\mathbf{u}}{\text{)}}}&{\mathbf{x}} \\ 
  {{{\mathbf{x}}^ * }}&t 
\end{array}} \right]} \right\|_F^2, \\ 
\end{gathered} 
\end{equation}
where $\bf{\Lambda}$ is the Lagrange multiplier, and $\rho$ is the regularization parameter. The ADMM algorithm proceeds with the iterations as follows:
\begin{equation} \label{eq:13}
\left( {{{\bf{u}}^{(i + 1)}},{{\bf{x}}^{(i + 1)}},{t^{(i + 1)}}} \right) = \arg \mathop {\min }\limits_{{\bf{u}},{\bf{x}},t} \mathcal L\left( {{\bf{u}},{\bf{x}},t,{{\bf{Z}}^{(i)}},{{\bf{\Lambda }}^{(i)}}} \right),
\end{equation}

\begin{equation} \label{eq:14}
    {{\mathbf{Z}}^{(i + 1)}} = \arg \mathop {\min }\limits_{{\mathbf{Z}} \succeq 0} \mathcal{L}\left( {{{\mathbf{u}}^{(i + 1)}},{{\mathbf{x}}^{(i + 1)}},{t^{(i + 1)}}, {\mathbf{Z}},{{\mathbf{\Lambda }}^{(i)}}} \right),
\end{equation}

\begin{equation} \label{eq:15}
    {{\mathbf{\Lambda }}^{(i + 1)}} = {{\mathbf{\Lambda }}^{(i)}} + \rho \left( {{{\mathbf{Z}}^{(i + 1)}} - \left[ {\begin{array}{*{20}{c}}
  {{{T(}}{{\mathbf{u}}^{(i + 1)}}{\text{)}}}&{{{\mathbf{x}}^{(i + 1)}}} \\ 
  {{{\mathbf{x}}^{ * (i + 1)}}}&{{t^{(i + 1)}}} 
\end{array}} \right]} \right).
\end{equation}

The updates of $t$, $\bf{u}$ and $\bf{x}$ are in closed form:
\begin{equation} \label{eq:16}
    {{\mathbf{u}}^{(i + 1)}} = \frac{1}{{\mathcal{P}({\mathbf{I}})}}\left( {\mathcal{P}\left( {{\mathbf{Z}}_{\mathbf{u}}^{(i)} + \frac{1}{\rho }{\mathbf{\Lambda }}_{\mathbf{u}}^{(i)}} \right) - \frac{\tau }{{2\rho }}{{\mathbf{e}}_{\mathbf{1}}}} \right),
\end{equation}

\begin{equation} \label{eq:17}
    {{\mathbf{x}}^{(i + 1)}} = {\left( {\frac{1}{2}{{\mathbf{D}}^ * }{\mathbf{D}} + \rho {\mathbf{I}}} \right)^{ - 1}}\left( {\rho {\mathbf{Z}}_{\mathbf{x}}^{(i)} + {\mathbf{\Lambda }}_{\mathbf{x}}^{(i)} + \frac{1}{2}{{\mathbf{D}}^ * }{\mathbf{x}}} \right),
\end{equation}

\begin{equation} \label{eq:18}
    {t^{(i + 1)}} = \frac{1}{\rho }{\mathbf{\Lambda }}_t^{(i)} + {\mathbf{Z}}_t^{(i)} - {\mathbf{I}}\frac{\tau }{{2\rho }},
\end{equation}
where the first entry of vector $\bf{e}_1$  equals 1 and the rest equal 0. $\mathcal{P}( \cdot )$ is a pseudo-inverse operator. The update step of $\bf{Z}$ is to project onto the SDP cone, which is:
\begin{equation} \label{eq:19}
    {{\mathbf{Z}}^{(i + 1)}} = \arg \mathop {\min }\limits_{{\mathbf{Z}} \succeq 0} \left\| {{{\mathbf{Z}}^{(i)}} - \left[ {\begin{array}{*{20}{c}}
  {{\rm{Toep(}}{{\mathbf{u}}^{(i + 1)}}{\text{)}}}&{{{\mathbf{x}}^{(i + 1)}}} \\ 
  {{{\mathbf{x}}^{ * (i + 1)}}}&{{t^{(i + 1)}}} 
\end{array}} \right] + \frac{1}{\rho }{{\mathbf{\Lambda }}^{(i)}}} \right\|_F^2,
\end{equation}
where the partitions are denoted as:
\begin{equation} \label{eq:20}
    {\mathbf{Z}} = \left[ {\begin{array}{*{20}{c}}
  {{{\mathbf{Z}}_{\mathbf{u}}}}&{{{\mathbf{Z}}_{\mathbf{x}}}} \\ 
  {{\mathbf{Z}}_{\mathbf{x}}^ * }&{{{\mathbf{Z}}_t}} 
\end{array}} \right],{\mathbf{\Lambda }} = \left[ {\begin{array}{*{20}{c}}
  {{{\mathbf{\Lambda }}_{\mathbf{u}}}}&{{{\mathbf{\Lambda }}_{\mathbf{x}}}} \\ 
  {{\mathbf{\Lambda }}_{\mathbf{x}}^ * }&{{{\mathbf{\Lambda }}_t}} 
\end{array}} \right].
\end{equation}

In summary, the pseudocode of the solving algorithm for AST-STF is shown in Algorithm \ref{alg:1}.

\begin{algorithm}[ht]
\caption{The ADMM solving algorithm for AST-STF (\ref{eq:11}).}
\label{alg:1}
\begin{algorithmic}[1] %
\REQUIRE ~~\\ %
    Measured signal ${\mathbf{y}} \in {\mathbb{R}^N}$, de-window matrix ${\mathbf{D}} \in {\mathbb{R}^{N \times (LW)}}$, regularization parameter $\tau$, and penalty parameter $\rho$;
\ENSURE ${{\mathbf{\hat u}}^{(i + 1)}}$, ${{\mathbf{\hat x}}^{(i + 1)}}$; 
    \STATE Calculate: ${\left( {\frac{1}{2}{{\mathbf{D}}^ * }{\mathbf{D}} + \rho {\mathbf{I}}} \right)^{ - 1}}$;
    \STATE Initialize: ${{\mathbf{Z}}^{\left( 0 \right)}}$, ${{\mathbf{\Lambda }}^{\left( 0 \right)}}$;
    \FOR{$i=1:M_{max}$}
    \STATE Update $\bf{u}^{(i+1)}$, $\bf{x}^{(i+1)}$, $t^{(i+1)}$, $\bf{Z}^{(i+1)}$, $\bf{\Lambda}^{(i+1)}$;
    \ENDFOR
\RETURN  ${{\mathbf{\hat u}}^{(i + 1)}}$, ${{\mathbf{\hat x}}^{(i + 1)}}$.
\end{algorithmic}
\end{algorithm}

\subsection{Localizing support frequencies via dual polynomial}
The dual problem of (\ref{eq:9}) is: 
\begin{equation} \label{eq:21}
\begin{gathered}
  \mathop {\max }\limits_{{\mathbf{z}} \in {\mathbb{F}^N}} \frac{1}{2}\left( {\left\| {\mathbf{y}} \right\|_2^2 - \left\| {{\mathbf{y}} - \tau {\mathbf{z}}} \right\|_2^2} \right), \hfill \\
  {\text{s}}{\text{.t}}{\text{. }}\left\| {{{\mathbf{D}}^ * }{\mathbf{z}}} \right\|_\mathcal{A}^ *  \leqslant 1, \hfill \\ 
\end{gathered} 
\end{equation}
where dual atomic norm $\left\|  \cdot  \right\|_\mathcal{A}^ * $ denotes the support function. According to the Lemma 2 in \cite{bhaskar_atomic_2013}, dual problem (\ref{eq:21}) guarantees the unique optimal solution ${\mathbf{\hat z}} \in {\mathbb{C}^{LW}}$ because of its strong concavity, which also illustrates zero dual gap between ${\mathbf{\hat x}}$ and $\mathbf{\hat z}$, i.e., the primal and dual solutions. Once ${\mathbf{\hat x}}$ is solved, its dual solution ${\mathbf{\hat z}} = {{\mathbf{D}}^ * }\left( {{\mathbf{y}} - {\mathbf{Dx}}} \right)$ can be obtained.

According to the dual certificate support from Corollary 1 in \cite{bhaskar_atomic_2013}, the dual solution ${{\mathbf{\hat z}}_w} \in {\mathbb{C}^L}$ of the $w$-th windowed segment is the optimum satisfying:
\begin{equation} \label{eq:22}
    \begin{gathered}
  \left| {\left\langle {{{{\mathbf{\hat z}}}_w},{\mathbf{a_i}}} \right\rangle } \right| = \tau ,{\forall \mathbf{a_i}} \in S, \hfill \\
  \left| {\left\langle {{{{\mathbf{\hat z}}}_w},{\mathbf{a_i}}} \right\rangle } \right| < \tau ,{\forall \mathbf{a_i}} \notin S, \hfill \\ 
\end{gathered} 
\end{equation}
where $S\subset\mathcal{A}$ is the support set. Therefore, we can localize the support frequencies in each windowed segment by calculating the modulus of the dual polynomial ${H_w}\left( f \right) = \left\| {{{{\mathbf{\hat z}}}_w}} \right\|_\mathcal{A}^ * $, which can be efficiently solved by Fast Fourier transform (FFT) with fine grids to achieve excellent frequency resolution without considering spectral leakage. 

\begin{figure}[b] 
\centerline{\includegraphics[width=0.5\columnwidth]{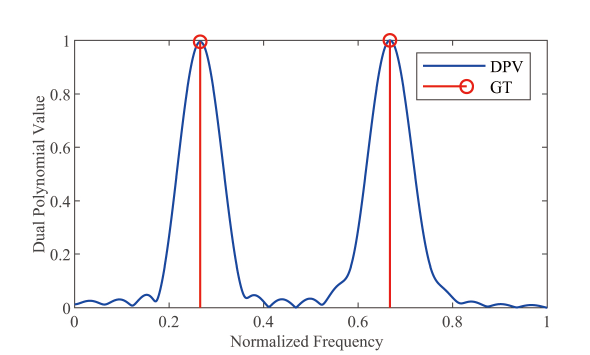}}
\caption{Dual polynomial value (DPV, in blue) and its ground truth (GT, in red) support frequencies of a windowed segment.}
\label{fig:2}
\end{figure}

In the aftermath, the least square method can be used to calculate the amplitude of each support frequency. Fig. \ref{fig:2} shows the dual polynomial value (DPV) of a signal segment truncated by a window, and its localized support frequencies correspond to the values closest to the maximum, which are the dualizing and localizing operations described above.

\section{Simulation}
\label{sec:simulation}
We present a simulation study in this section to analyze the performance of AST-STF. A noise-free signal is firstly analyzed to demonstrate the capability of AST-STF in enhancing TF energy concentration. Subsequently, the robustness of AST-STF against noise is investigated.

\subsection{Enhancement of TF energy concentration}
\begin{figure}[t]
\centering
  \begin{minipage}[b]{0.35\linewidth}
    \centering
    \includegraphics[width=\linewidth]{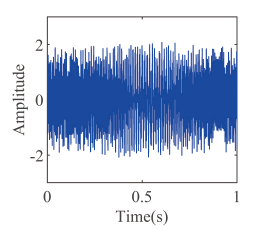}
    \subcaption{Waveform}
  \end{minipage}
  \hspace{0.02\linewidth}
  \begin{minipage}[b]{0.35\linewidth}
    \centering
    \includegraphics[width=\linewidth]{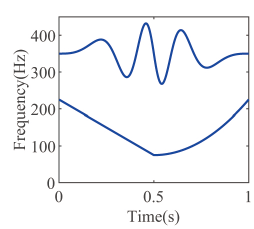}
    \subcaption{Instant frequency}
  \end{minipage}
  \caption{Waveform and IF of the generated signal $x(t)$.}
  \label{fig:3}
\end{figure}

To illustrate the performance of AST-TF method in enhancing TF energy concentration, we generate a multi-component signal with slowly and rapidly varying IF structures \cite{wangReassignmentenableReweightedSparse2023, tongRidgeAwareWeightedSparse2021}:
\begin{equation} \label{eq:23}
\begin{array}{l}
{x_1}(t) = \left( {1 - 0.5\cos (2\pi t)} \right)\cos \left( {700\pi t - 15\cos \left( {12\pi {{\left( {{\rm{0}}.{\rm{5}} - \left| {t - 0.5} \right|} \right)}^2}} \right)} \right), \\
{x_2}(t) = \left\{ {\begin{array}{*{20}{c}}
\begin{array}{l}
\cos \left( {450{\rm{\pi }}t - 300{\rm{\pi }}{t^2}} \right),\\
\cos \left( {450{\rm{\pi }}t - 600{\rm{\pi }}{t^2} + 400{\rm{\pi }}{t^3} + {\rm{\pi }}} \right),
\end{array}&{\begin{array}{*{20}{c}}
{0 \le t < 0.5},\\
{0.5 \le t < 1},
\end{array}}
\end{array}} \right.\\
x(t) = {x_1}(t) + {x_2}(t),
\end{array}
\end{equation}
where $x_1(t)$ is a frequency and amplitude modulated signal with rapidly varying IF and amplitude. $x_2(t)$ is composed of a linear chirp and a quadratic chirp. The signal is sampled at a frequency of $f_s=1024Hz$, with a total length of $N=1024$. The waveform and IF of the generated signal are presented in Fig. \ref{fig:3}.

\begin{figure}[ht]
\centering
  \begin{minipage}[b]{0.3\linewidth}
    \centering
    \includegraphics[width=\linewidth]{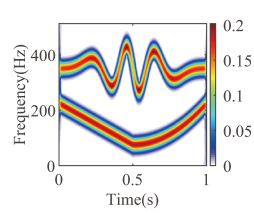}
    \subcaption{STFT}
  \end{minipage}
  \hspace{0.02\linewidth}  
  \begin{minipage}[b]{0.3\linewidth}
    \centering
    \includegraphics[width=\linewidth]{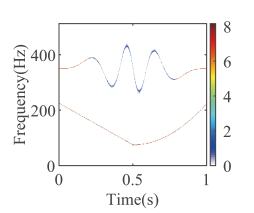}
    \subcaption{Reassignment}
  \end{minipage}
  \\[0.1cm]  
  \begin{minipage}[b]{0.3\linewidth}
    \centering
    \includegraphics[width=\linewidth]{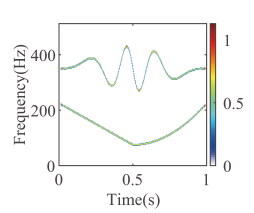}
    \subcaption{STFT-based sparse representation}
  \end{minipage}
  \hspace{0.02\linewidth}  %
  \begin{minipage}[b]{0.3\linewidth}
    \centering
    \includegraphics[width=\linewidth]{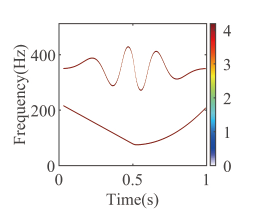}
    \subcaption{AST-STF}
  \end{minipage}
  \caption{TF distributions of the noise-free signal $x(t)$.}
  \label{fig:4}
\end{figure}

We compare the results of three TF representation methods (STFT, reassignment, STFT-based sparse representation method) with our proposed AST-STF, the respective TF distributions are shown in Fig. \ref{fig:4}. Since the performances of reassignment and AST-TF are influenced by the selected window, we conduct the experiments with the window lengths that correspond to the most effective results.

\begin{figure}[htb]
\centering
  \begin{minipage}[b]{0.3\linewidth}
    \centering
    \includegraphics[width=\linewidth]{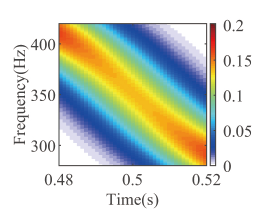}
    \subcaption{STFT}
  \end{minipage}
  \hspace{0.02\linewidth}  
  \begin{minipage}[b]{0.3\linewidth}
    \centering
    \includegraphics[width=\linewidth]{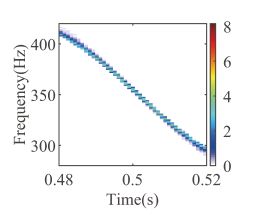}
    \subcaption{Reassignment}
  \end{minipage}
  \\[0.1cm]  
    \begin{minipage}[b]{0.3\linewidth}
    \centering
    \includegraphics[width=\linewidth]{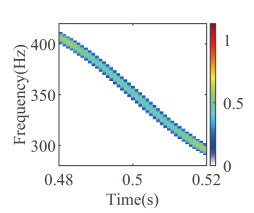}
    \subcaption{STFT-based sparse representation}
  \end{minipage}
  \hspace{0.02\linewidth}  
  \begin{minipage}[b]{0.3\linewidth}
    \centering
    \includegraphics[width=\linewidth]{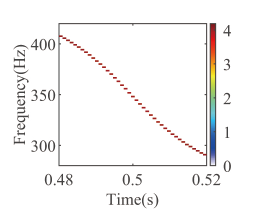}
    \subcaption{AST-STF}
  \end{minipage}
  \caption{Zoomed-in view of the component $x_1(t)$ in Fig. \ref{fig:4} .}
  \label{fig:5}
\end{figure}

The results can be illustrated clearly in Fig. \ref{fig:5}, with a zoomed-in view of the region [0.48, 0.52]s $\times$ [280, 420]Hz for the component $x_1(t)$. From a qualitative perspective, compared to the conventional methods, it turns out that the proposed AST-STF proves to be significantly more effective in TF localization and energy concentration.

To analyze the energy concentration quantitatively, we employ the third-order Rényi entropy \cite{baraniuk_measuring_2001,aviyente_minimum_2005}, which is defined as: 
\begin{equation} \label{eq:24}
    R_C^\alpha  = \frac{1}{{1 - \alpha }}{\log _2}\sum\limits_n {\sum\limits_k {{{\left( {\frac{{C\left[ {n,k} \right]}}{{\sum\limits_n {\sum\limits_k {C\left[ {n,k} \right]} } }}} \right)}^\alpha }} } ,\alpha  = 3.
\end{equation}

The value of Rényi entropy decreases as the TF energy concentration improves. The corresponding values for different TF representation methods are shown in TABLE \ref{tab:1}. It is evident that for the noise-free signal, AST-STF can achieve the lowest Rényi entropy, indicating superior TF energy concentration.

\clearpage

\begin{table}[width=.5\linewidth,cols=2,pos=t]
\caption{Rényi entorpy of different TF representations.}
\begin{tabular*}{\tblwidth}{@{\hskip 5pt} LC @{\hskip 5pt} }
\toprule
\textbf{TF Representations} & \multicolumn{1}{l}{\textbf{Rényi Entorpy}} \\ 
\midrule
STFT                        & 6.862                                      \\
Reassignment                & 1.452                                      \\
STFT-based sparse representation & 2.984                                      \\
AST-STF                     & \textbf{0.161}                                     \\ 
\bottomrule
\end{tabular*}
\label{tab:1}
\end{table}

\subsection{Denoising using AST-STF }

To evaluate the denoising performance of AST-STF, we add Gaussian white noise with signal-to-noise ratio SNR = 5 dB (standard deviation $\sigma=0.5614$) to the signal $x(t)$.

\begin{figure}[t]
\centering
  \begin{minipage}[b]{0.3\linewidth}
    \centering
    \includegraphics[width=\linewidth]{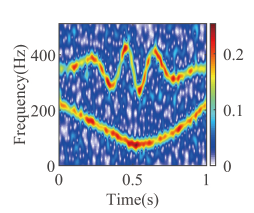}
    \subcaption{STFT}
  \end{minipage}
  \hspace{0.02\linewidth}  
  \begin{minipage}[b]{0.3\linewidth}
    \centering
    \includegraphics[width=\linewidth]{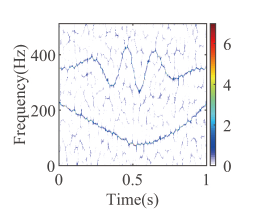}
    \subcaption{Reassignment}
  \end{minipage}
  \\[0.1cm]
    \begin{minipage}[b]{0.3\linewidth}
    \centering
    \includegraphics[width=\linewidth]{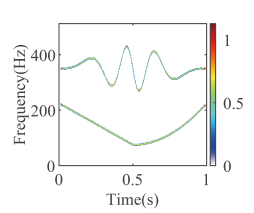}
    \subcaption{STFT-based sparse representation}
  \end{minipage}
  \hspace{0.02\linewidth}  
  \begin{minipage}[b]{0.3\linewidth}
    \centering
    \includegraphics[width=\linewidth]{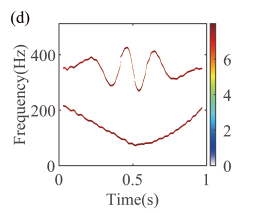}
    \subcaption{AST-STF}
  \end{minipage}
  \caption{TF distributions of the noisy signal.}
  \label{fig:6}
\end{figure}

The results of TF representations are shown in Fig. \ref{fig:6}. The minimum root-mean-squared error (RMSE) of AST-STF is 0.209, which is smaller than the other methods, as listed in TABLE \ref{tab:2}. It means that the proposed AST-STF outperforms the others in signal denoising. In TABLE \ref{tab:2}. we also assess the Rényi entropy for all the TF representations, which demonstrates that AST-STF method retains better TF energy concentration (i.e., has lower Rényi entropy) in the presence of noise.

\begin{table}[width=.5\linewidth,cols=3,pos=t]
\centering
\caption{Rényi entorpy and RMSE of TF representations.}
\begin{tabular}{lcc}
\toprule
\textbf{TF Representations} & \multicolumn{1}{l}{\textbf{Renyi Entorpy}} & \multicolumn{1}{l}{\textbf{RMSE}} \\ 
\midrule
STFT                        & 7.884                                      & \textbackslash{}                  \\
Reassignment                & 4.038                                      & \textbackslash{}                  \\
STFT-based sparse representation & 3.997                                      & 0.234                             \\
AST-STF                     & \textbf{0.106}                            & \textbf{0.209}                    \\ 
\bottomrule
\end{tabular}
\label{tab:2}
\end{table}


\section{Application}
\label{sec:application}
\begin{figure}[t] 
\centerline{\includegraphics[width=0.5\columnwidth]{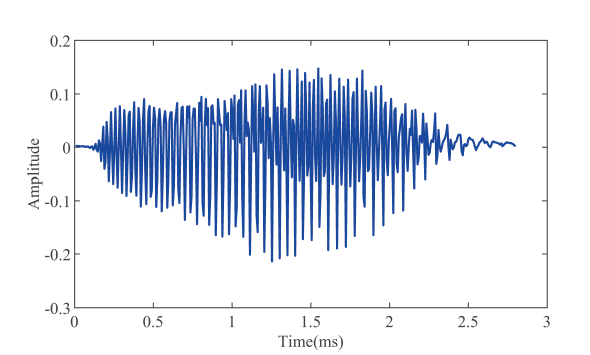}}
\caption{Waveform of the bat echolocation signal.}
\label{fig:7}
\end{figure}

This section presents the benchmark practical application on a bat echolocation signal to verify the effectiveness of the proposed AST-STF method. The sampling period and samples are 2.7 microseconds and 400 samples, respectively. The waveform of the signal is shown in Fig. \ref{fig:7}.

\begin{figure}[t]
\centering
  \begin{minipage}[b]{0.3\linewidth}
    \centering
    \includegraphics[width=\linewidth]{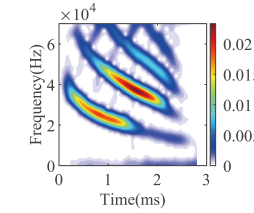}
    \subcaption{STFT}
  \end{minipage}
  \hspace{0.02\linewidth}  
  \begin{minipage}[b]{0.3\linewidth}
    \centering
    \includegraphics[width=\linewidth]{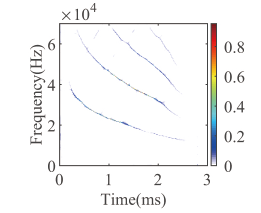}
    \subcaption{Reassignment}
  \end{minipage}
  \\[0.1cm]
    \begin{minipage}[b]{0.3\linewidth}
    \centering
    \includegraphics[width=\linewidth]{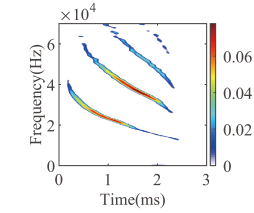}
    \subcaption{STFT-based sparse representation}
  \end{minipage}
  \hspace{0.02\linewidth}  
  \begin{minipage}[b]{0.3\linewidth}
    \centering
    \includegraphics[width=\linewidth]{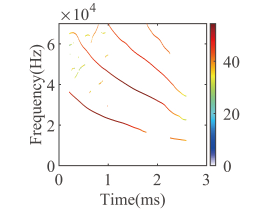}
    \subcaption{AST-STF}
  \end{minipage}
  \caption{TF distributions of the bat echolocation signal.}
  \label{fig:8}
\end{figure}

\begin{figure}[t]
\centering
  \begin{minipage}[b]{0.3\linewidth}
    \centering
    \includegraphics[width=\linewidth]{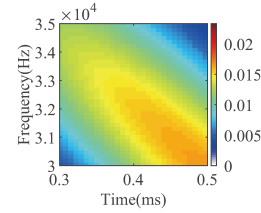}
    \subcaption{STFT}
  \end{minipage}
  \hspace{0.02\linewidth}  
  \begin{minipage}[b]{0.3\linewidth}
    \centering
    \includegraphics[width=\linewidth]{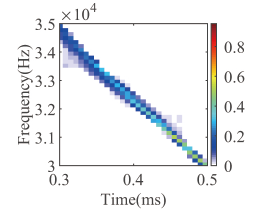}
    \subcaption{Reassignment}
  \end{minipage}
  \\[0.1cm]
    \begin{minipage}[b]{0.3\linewidth}
    \centering
    \includegraphics[width=\linewidth]{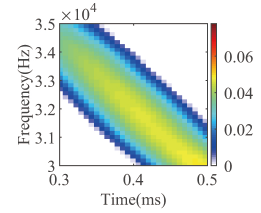}
    \subcaption{STFT-based sparse representation}
  \end{minipage}
  \hspace{0.02\linewidth}  
  \begin{minipage}[b]{0.3\linewidth}
    \centering
    \includegraphics[width=\linewidth]{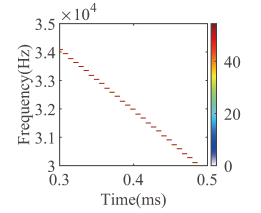}
    \subcaption{AST-STF}
  \end{minipage}
  \caption{Zoomed-in view in Fig. \ref{fig:8}.}
  \label{fig:9}
\end{figure}

The TF distributions of AST-STF and other conventional TF representations are shown in Fig. \ref{fig:8}. Notably, AST-STF method exhibits much greater strength in extracting and enhancing weak features of the signal. Moreover, the resulting representation is considerably sparser than the others, particularly for the rapidly varying IF component. The results can be more clearly observed in the zoomed-in view of the region [0.3, 0.5]s $\times$ [30, 35] kHz, as shown in Fig. \ref{fig:9}.

To quantitatively demonstrate the performance of AST-STF in enhancing TF energy concentration, we compute the third-order Rényi entropy of different TF representation methods, as listed in TABLE \ref{tab:3}. The results clearly indicate that the proposed AST-STF significantly enhances TF energy concentration.

\begin{table}[width=.5\linewidth,cols=2,pos=t]
\centering
\caption{Rényi entorpy of TF representations.}
\begin{tabular*}{\tblwidth}{@{\hskip 5pt} LC @{\hskip 5pt} }
\toprule
\textbf{TF Representations} & \multicolumn{1}{l}{\textbf{Rényi Entorpy}} \\ 
\midrule
STFT                        & 16.997                                      \\
Reassignment                & 12.245                                      \\
STFT-based sparse representation & 13.067                                      \\
AST-STF                     & \textbf{4.634}                                     \\ 
\bottomrule
\end{tabular*}
\label{tab:3}
\end{table}

\section{Conclusions}
\label{sec:conclusion}

In this paper, we propose a novel method called AST-STF, which serves as an effective TFA tool for analyzing non-stationary signals with weak and rapidly varying IF components. The primary advantage of AST-STF, compared to conventional TF representations, is its capability to ensure extraction of IF while maintaining near maximal-sparsity in TF domain, i.e., the outstanding TF energy concentration performance even as compared to the state-of-the-art reassignment method. The main reason is that the proposed TF representation method based on AST can accurately localize support frequencies in the continuous atomic domain, even within relatively short signal segment truncated by window, under which the guarantee of strong duality between the primal and dual problems.
The efficacy of the proposed AST-STF is demonstrated through numerical simulation and benchmark practical application. Comparative analysis shows that the proposed method outperforms STFT, reassignment and STFT-based sparse representation method in terms of TF energy concentration enhancing and signal denoising. 

Several extensions of the study are of interest. Despite the ADMM algorithm has substantially reduced computational expenses in solving AST-STF analysis model, the SDP still poses limitations on applicability in large-scale signal processing, which requires further study of accelerating algorithm. Moreover, the smoothness improvement of IF under short window and the reconstruction performance are still worth studying. Finally, the presence of sophisticated  noise  interference in signals makes the selection of regularization parameter challenging, which requires further investigating.

\printcredits

\section{Declaration of competing interest}
The authors declare that they have no known competing financial interests or personal relationships that could have appeared to influence the work reported in this paper.

\section{Data availability}
Data will be made available on request.

\section{Acknowledgments}
This work was supported by the National Natural Science Foundation of China under Grant 52105121, 92060302, and 52122504, and the National Major Science and Technology Projects of China under Grant J2019-IV-0018.



\end{document}